# What does Newcomb's paradox teach us?

David H. Wolpert[*], Gregory Benford[†]

February 26, 2010

**Abstract**

In Newcomb's paradox you choose to receive either the contents of a particular closed box, or the contents of both that closed box and another one. Before you choose, a prediction algorithm deduces your choice, and fills the two boxes based on that deduction. Newcomb's paradox is that game theory appears to provide two conflicting recommendations for what choice you should make in this scenario. We analyze Newcomb's paradox using a recent extension of game theory in which the players set conditional probability distributions in a Bayes net. We show that the two game theory recommendations in Newcomb's scenario have different presumptions for what Bayes net relates your choice and the algorithm's prediction. We resolve the paradox by proving that these two Bayes nets are incompatible. We also show that the accuracy of the algorithm's prediction, the focus of much previous work, is irrelevant. In addition we show that Newcomb's scenario only provides a contradiction between game theory's expected utility and dominance principles if one is sloppy in specifying the underlying Bayes net. We also show that Newcomb's paradox is time-reversal invariant; both the paradox and its resolution are unchanged if the algorithm makes its 'prediction' *after* you make your choice rather than before.

[*]NASA Ames Research Center, MS 269-1, Moffett Field, CA 94035-1000, (650) 604-3362 (V), (650) 604-3594 (F), david.h.wolpert.nasa.gov

[†]Physics and Astronomy Department, University of California, Ivine, CA 92692

# 1 Introduction

## 1.1 Background

Suppose you meet a Wise being (*W*) who tells you it has put $1,000 in box A, and either $1 million or nothing in box B. This being tells you to either take the contents of box B only, or to take the contents of both A and B. Suppose further that the being had put the $1 million in box B only if a prediction algorithm designed by the being had said that you would take only B. If the algorithm had predicted you would take both boxes, then the being put nothing in box B.

Presume that due to determinism, there exists a perfectly accurate prediction algorithm. Assume that *W* uses that algorithm to make its prediction, and after that you make your choice. However when you make your choice, you don't know what that prediction is. What choice should you make?

In Table 1 we present this question as a game theory matrix involving *W*'s prediction and your choice. Two seemingly logical answers contradict each other. The Realist answer is that you should take both boxes, because your choice occurs after *W* has already made its prediction, and since you have free will, you're free to make whatever choice you want, independent of that prediction that *W* made. More precisely, if *W* predicted you would take A along with B, then taking both gives you $1,000 rather than nothing. If instead *W* predicted you would take only B, then taking both boxes yields $1,001,000, which again is $1000 better than taking only B. (In the language of game theory, choosing both A and B is the dominant strategy [1, 2, 3].) In contrast, the Fearful answer is that *W* designed a prediction algorithm whose answer will match what you do. So you can get $1,000 by taking both boxes or get $1 million by taking only box B. Therefore you should take only B.

This is the conventional formulation of Newcomb's Paradox, a famous logical riddle stated by William Newcomb in 1960 [4, 5, 6, 7, 8, 9, 10, 11, 12, 13, 14]. Newcomb never published the paradox, but had long conversations about it with with philosophers and physicists such as Robert Nozick and Martin Kruskal, along with Scientific American's Martin Gardner. Gardner said after his second Scientific American column on Newcomb's paradox appeared that it generated more mail than any other column.

One of us (Benford) worked with Newcomb, publishing several papers together. We often discussed the paradox, which Newcomb invented to test his own ideas. Newcomb said that he would just take B; why fight a God-like being? However Nozick said, "To almost everyone, it is perfectly clear and obvious what should be done. The difficulty is that these people seem to divide almost evenly on the problem, with large numbers thinking that the opposing half is just being silly" [4].

Nozick also pointed out that two accepted principles of game theory appear to conflict in Newcomb's problem. The expected-utility principle, considering the probability of each outcome, says you should take box B only. But the dominance principle argues that if one strategy is always better than the other strategies no matter what other players do, then you should pick that strategy. No matter what box B contains, you are $1000 richer if you take both boxes than if you take B only. So the dominance principle says you should take both boxes.

Is there really a contradiction? Some philosophers argue that a perfect predictor implies a time machine, since with such a machine causality is reversed, i.e., for predictions made in the present to be perfectly determined by events in the future means that the future causes past events.[1]

But Nozick stated the problem specifically to exclude backward causation (and so time travel), because his formulation demands only that the predictions be of high accuracy, not certain. So this line of reasoning cannot resolve the issue.

## 1.2 Game theory over Bayes nets

Central to Newcomb's scenario is a prediction process, and its (in)fallibility. Recent work has revealed deep formal connections between prediction and observation. Amongst other things, this work proves that any given prediction algorithm must fail on at least one prediction task [16, 17]. That particular result does not resolve Newcomb's paradox. However the proof of that result requires an extension of conventional game theory. And as we show below, that extension of game theory provides a way to resolve Newcomb's paradox.

In conventional game theory there are several 'players', each with their own preferences over the values of an underlying set of *game variables*, $\{X_j\}$. Every player has their own *strategy set*, where each strategy is itself a set, of conditional probability distributions relating some of the game variables $\{X_j\}$. To play the game, the players all independently choose a strategy (i.e., choose a set of distributions) from their respective strategy sets. The strategies sets are carefully designed so that every such joint strategy of the players uniquely specifies a legal

---

[1]Interestingly, near when Newcomb devised the paradox, he also coauthored a paper proving that a tachyonic time machine could not be reinterpreted in a way that precludes such paradoxes [15]. The issues of time travel and Newcomb-style paradoxes are intertwined.

joint probability distribution relating all the game's variables [1, 2, 3].

A richer mathematics arises if we expand the strategy sets of the players, so that some joint strategies would violate the laws of probability, and therefore are impossible. Formally, in this extension of conventional game theory, multiple different Bayes nets relating the game variables are specified, and each player sets conditional distributions at some nodes in each of those Bayes nets. So a player's strategy now specifies conditional distributions in all of the Bayes nets at once. (In contrast, conventional game theory implicitly only allows a single Bayes net.) This means that some joint player strategies may be impossible, in that they result in each Bayes net having a different joint distribution over the game variables. Whenever the joint distributions of two Bayes nets are inconsistent this way, some set of the conditional distributions in one of the two nets will conflict with some set of distributions in the other Bayes net.

It is exactly such inconsistent joint distributions that underlie the fallibility of prediction proven in [16, 17]. As we show below, this possibility of inconsistency also arises when one properly formalizes Newcomb's scenario in terms of games. In turn, this possibility of inconsistency resolves Newcomb's paradox. Loosely speaking, Newcomb's scenario involves two Bayes nets, one corresponding to Realist's reasoning, and one corresponding to Fearful's reasoning. One can choose one Bayes net or the other, but to avoid inconsistency, one cannot choose both. In short, Realist and Fearful implicitly formalize Newcomb's scenario in different, and incompatible ways.

In the next section, we first show how to express the reasoning of Realist and of Fearful in terms of strategy spaces over two different Bayes nets. We then show that those two Bayes nets and associated strategy spaces cannot be combined without raising the possibility of inconsistent joint distributions between the Bayes nets. This resolves Newcomb's paradox, without relying on mathematizations of concepts that are vague and/or controversial (e.g., 'free will', 'causality'). In addition, our resolution shows that there is no conflict in Newcomb's scenario between the dominance principle of game theory and the expected utility principle. Once one takes care to specify the underlying Bayes net game, those principles are completely consistent with one another.

We end by discussing the implications of our analysis. In particular, our analysis formally establishes the (un)importance of $W$'s predictive accuracy. It also provides an illustrative way to look at Newcomb's scenario by running that scenario backwards in time.

# 2 Resolving Newcomb's paradox with extended game theory

There are two players in Newcomb's paradox: you, and the wise being $W$. In addition, there are two game variables that are central to the paradox: $W$'s prediction, $g$, and the choice you actually make, $y$. So the player strategies will involve the joint probability distribution relating those two variables. Since there are only two variables, there are only two ways to decompose that joint probability distribution into a Bayes net. These two decompositions turn out to correspond to the two recommendations for how to answer Newcomb's question, one matching the reasoning of Realist and one matching Fearful.

## 2.1 The first decomposition of the joint probability

The first way to decompose the joint probability is

$$P(y, g) = P(g \mid y)P(y) \tag{1}$$

(where we define the right-hand side to equal 0 for any $y$ such that $P(y) = 0$). Such a decomposition is known as a 'Bayes net' having two 'nodes' [18, 19]. The variable $y$, and the associated unconditioned distribution, $P(y)$, is identified with the first, 'parent' node. The variable $g$, and the associated conditional distribution, $P(g \mid y)$, is identified with the second, 'child' node. The stochastic dependence of $g$ on $y$ can be graphically illustrated by having a directed edge go from a node labeled $y$ to a node labeled $g$.

This Bayes net can be used to express Fearful's reasoning. Fearful interprets the statement that '$W$ designed a perfectly accurate prediction algorithm' to imply that $W$ has the power to set the conditional distribution in the child node of the Bayes net, $P(g \mid y)$, to anything it wants (for all $y$ such that $P(y) \neq 0$). More precisely, since the algorithm is 'perfectly accurate', Fearful presumes that $W$ chooses to set $P(g \mid y) = \delta_{g,y}$, the Kronecker delta function that equals 1 if $g = y$, zero otherwise. So Fearful presumes that there is nothing you can do that can affect the values of $P(g \mid y)$ (for all $y$ such that $P(y) \neq 0$). Instead, you get to choose the unconditioned distribution in the parent node of the Bayes net, $P(y)$. Intuitively, this choice constitutes your 'free will'.

Fearful's interpretation of Newcomb's paradox specifies what aspect of $P(y, g)$ you can choose, and what aspect is instead chosen by W. Those choices — $P(y)$ and $P(g \mid y)$, respectively — are the 'strategies' that you and $W$ choose. It is important to note that these strategies by you and $W$ do *not* directly specify the two variables $y$ and $g$. Rather the strategies you and $W$ choose specify two different distributions which, taken together, specify the full joint distribution over $y$ and $g$ [19]. This kind of strategy contrasts with the kind considered in decision theory [20] or causal nets [18], where the strategies are direct specifications of the variables (which here are $g$ and $y$).

In game theory, by the axioms of VonNeumann Morgenstern utility theory, your task is to choose the strategy that maximizes your expected payoff under the associated joint distribution.[2] For Fearful, this means choosing the $P(y)$ that maximizes your expected payoff under the $P(y, g)$ associated with that choice. Given Fearful's presumption that the Bayes net of Eq. 1 underlies the game with $P(g \mid y) = \delta_{g,y}$, and that your strategy is to set the distribution at the first node, for you to maximize expected payoff your strategy should be $P(y) = \delta_{y,B}$. In other words, you should choose $B$ with probability 1. Your doing so results in the joint distribution $P(y, g) = \delta_{g,y}\delta_{y,B} = \delta_{g,B}\delta_{y,B}$, with payoff $1,000,000$. This is the formal justification of Fearful's recommendation.[3]

## 2.2 The second decomposition of the joint probability

The second way to decompose the joint probability is

$$P(y, g) = P(y \mid g)P(g) \qquad (2)$$

(where we define the right-hand side to equal 0 for any $g$ such that $P(g) = 0$). In the Bayes net of Eq. 2, the unconditioned distribution identified with the parent node is $P(g)$, and the conditioned distribution identified with the child node is $P(y \mid g)$. This Bayes net can be used to express Realist's reasoning. Realist interprets the statements that 'your choice occurs after $W$ has already made its prediction' and 'when you have to make your choice, you don't know what that prediction is' to mean that you can choose any distribution $h(y)$ and then set $P(y \mid g)$ to equal $h(y)$ (for all $g$ such that $P(g) \neq 0$). This is how Realist interprets your having 'free will'. (Note that this is a different interpretation of 'free will" from the one made by Fearful.) Under this interpretation, $W$ has no power to affect $P(y \mid g)$. Rather $W$ gets to set the parent node in the Bayes net, $P(g)$. For Realist, this is the distribution that you cannot affect. (In contrast, in Fearful's reasoning, you set a non-conditional distribution, and it is the conditional distribution that you cannot affect.)

---

[2]For simplicity, we make the common assumption that utility is a linear function of payoff [21]. So maximizing expected utility is the same as maximizing expected payoff.

[3]In Fearful's Bayes net, $y$ 'causally influences $g$', to use the language of causal nets [18]. To cast this kind of causal relationship in terms of conventional game theory, we would have to replace the single-stage game in Table 1 with a two-stage game in which you first set $y$, and *then* $W$ strategies, having observed $y$. This two-stage game is incompatible with Newcomb's stipulation that $W$ strategies before you do, not after. This is one of the reasons why it is necessary to use extended game theory rather than conventional game theory to formalize Fearful's reasoning.

Realist's interpretation of Newcomb's paradox specifies what it is you can set concerning $P(y, g)$, and what is set by W. Just like under Fearful's reasoning, under Realist's reasoning the 'strategies' you and $W$ choose do not directly specify the variables $g$ and $y$. Rather the strategies of you and $W$ specify two distributions which, taken together, specify the full joint distribution. As before, your task is to choose your strategy — which now is $h(y)$ — to maximize your expected payoff under the associated $P(y, g)$. Given Realist's presumption that the Bayes net of Eq. 2 underlies the game and that you get to set $h$, you should choose $h(y) = P(y \mid g) = \delta_{y,AB}$, i.e., you should choose $AB$ with probability 1. Doing this results in the expected payoff $1,000 \, P(g = AB) + 1,001,000 \, P(g = B)$, which is your maximum expected payoff no matter what the values of $P(g = AB)$ and $P(g = B)$ are. This is the formal justification of Realist's recommendation.

Note that in Fearful's interpretation, your strategy is choosing a single real number, while W chooses two real numbers. In contrast, in Realist's interpretation, your strategy is still to choose a single real number, but now W also chooses only a single real number. The different interpretations correspond to different games.[4]

## 2.3 Combining the decompositions

The original statement of Newcomb's question is somewhat informal. In particular, as originally stated, and as commonly analyzed, Newcomb's question does not specify whether you are playing the game that Feaful thinks is being played, or the game that Realist thinks is being played. What happens if we try to formulate Newcomb's question to explicitly incorporate this lack of specificity? In other words, what happens if we try to merge the game that Fearful thinks is being played with the game that Realist thinks is being played?

---

[4]In Realist's Bayes net, given the associated restricted possible form of $P(y \mid g)$, there is no edge connecting the node for $g$ with the node for $y$. This means that $g$ and $y$ are 'causally independent', to use the language of causal nets [18]. This causal relationship is consistent with the single-stage game in Table 1, in contrast to the causal relationship of the game played under Fearful's interpretation, which requires a two-stage game.

Such a merging gives us an 'extended game' of the sort considered in [17], involving two Bayes net games. The Bayes nets of those two games are the two Bayes nets discussed in the preceding subsections. The strategy space of $W$ for the extended game is the set of all pairs of a conditional distribution $P(g \mid y)$ for the first (Fearful) Bayes net and a distribution $P(g)$ for the second (Realist) Bayes net. Your strategy space is the set of all pairs of a $P(y)$ for the first Bayes net, and a distribution $h(y)$ for the second Bayes net.

We will now show that there are several ways that this extended game can result in a joint distribution of the Fearful Bayes net game that contradicts the joint distribution of the Realist Bayes net game. Each of these ways that a contradiction can arise means that to make Newcomb's question fully formal, we cannot merge the Bayes net games; to avoid the possibility of a contradiction we must specify one or the other of the two games.

First, we will show that if $W$ sets $P(g \mid y)$ for Fearful's Bayes net appropriately, then some of your strategies $h(y)$ for Realist's Bayes net become impossible. In other words, for those strategy choices by $W$ and by you, the joint distribution over the game variables for Realists's Bayes net contradicts the joint distribution over the game variables for Fearful's Bayes net. (This is true for almost any $P(g \mid y)$ that $W$ might choose, and in particular is true even if $W$ does not predict perfectly.)

To begin, pick any $\alpha \in (1/2, 1]$, and presume that $P(g \mid y)$ is set by $W$ to be $\alpha \delta_{g,y} + (1-\alpha)(1-\delta_{g,y})$ (for all $y$ such that $P(y) \neq 0$). So for example, under Fearful's interpretation, where $W$ uses a perfectly error-free prediction algorithm, $\alpha = 1$. Given this presumption, the only way that $P(y \mid g)$ can be $g$-independent (for all $g$ such that $P(g) \neq 0$) is if it is one of the two delta functions, $\delta_{y,AB}$ or $\delta_{y,B}$. (See the appendix for a formal proof.) This contradicts Realist's interpretation of the game, under which you can set $P(y \mid g)$ to any $h(y)$ you desire.[5]

Similarly, if $P(g \mid y)$ is set by W to be $\delta_{g,y}$ for all $y$ such that $P(y) \neq 0$, as under Fearful's presumption, then your (Realist) choice of $h$ affects $P(g)$. In fact, your choice of $h$ fully specifies $P(g)$.[6] This contradicts Realist's presumption that it is $W$'s strategy that sets $P(g)$, independent of you.

Conversely, if you can set $P(y \mid g)$ to be an arbitrary $g$-independent distribution

---

[5]Note that of the two $\delta$ functions you can choose in this combined decomposition, it is better for you to choose $h(y) = \delta_{y,B}$, resulting in a payoff of $1,000,000$. So your optimal response to Newcomb's question for this variant is the same as if you were Fearful.

[6]For example, make Fearful's presumption be that $P(g \mid y) = \delta_{g,y}$ for all $y$ such that $P(y) \neq 0$. Given this, if you set $h(y) = \delta_{y,AB}$, then $P(g) = \delta_{g,AB}$, and if you set $h(y) = \delta_{y,B}$, then $P(g) = \delta_{g,B}$.

(as Realist presumes), then what you set it to may affect $P(g \mid y)$ (in violation of Fearful's presumption that $P(g \mid y)$ is set exclusively by W). In other words, if your having 'free will' means what it does to Realist, then you have the power to change the prediction accuracy of W (!). As an example, if you set $P(y = AB \mid g) = 3/4$ for all $g$'s such that $P(g) \neq 0$, then $P(g \mid y)$ cannot equal $\delta_{g,y}$.

The resolution of Newcomb's paradox is now immediate: You can be free to set $P(y)$ however you want, with $P(g \mid y)$ set by W, as Fearful presumes, *or*, as Realist presumes, you can be free to set $P(y \mid g)$ to whatever distribution $h(y)$ you want, with $P(g)$ set by W. It is not possible to play both games simultaneously.[7]

## 2.4 Discussion

It is important to emphasize that the impossibility of playing both games simultaneously arises for almost any error rate $\alpha$ in the $P(g \mid y)$ chosen by W, i.e., no matter how accurately W predicts. This means that the stipulation in Newcomb's paradox that W predicts perfectly is a red herring. (Interestingly, Newcomb himself did not insist on such perfect prediction in his formulation of the paradox, perhaps to avoid the time paradox problems.) The crucial impossibility implicit in Newcomb's question is the idea that at the same time you can arbitrarily specify 'your' distribution $P(y \mid g)$ and W can arbitrarily specify 'his" distribution $P(g \mid y)$. In fact, neither of you two can set your distribution without possibly affecting the other's distribution; you and W are inextricably coupled.

We also emphasize that vague concepts (e.g., 'free will') or controversial ones (e.g., 'causality') are not relevant to the resolution of Newcomb's paradox; it is not necessary to introduce mathematizations of those concepts to resolve Newcomb's paradox. The only mathematics needed is standard probability theory, together with the axioms of VonNeumann Morgenstern utility theory.

Another important contribution of our resolution is that it shows that Newcomb's scenario does not establish a conflict between game theory's dominance principle and its expected utility principle, as some have suggested. Indeed, as mentioned, we adopt the standard expected utility axioms of game theory throughout our analysis. However nowhere do we violate the dominance principle.

---

[7]In a variant of Newcomb's question, you first choose one of these two presumptions, and then set the associated distribution. If the pre-fixed distribution $P(g \mid y)$ arising in the first presumption is $\delta_{g,y}$, then your optimal responses depend on the pre-fixed distribution $P(g)$ arising in the second presumption— a distribution that is not specified in Newcomb's question. If $P(g)$ obeys $P(g = B) > .999$, then your optimal pair of choices are first to choose to set the distribution $P(y \mid g)$ to some $h(y)$, and then to set $h(y) = \delta_{y,AB}$. If this condition is not met, you should first choose to set $P(y)$, and then set it to $\delta_{y,AB}$.

The key behind our avoiding a conflict between those two principles is our taking care to specify what Bayes net underlies the game. Realist's reasoning *appears* to follow the dominance principle, and Fearful's to follow the principle of maximizing expected utility. (Hence the conflicting answers of Realist and Fearful appear to illustrate a conflict between those two principles.) However Realist is actually following the principle of maximizing expected utility *for that Bayes net game for which Realist's answer is correct*. In contrast, Realist's reasoning is an unjustified violation of that principle for the Bayes net game in which Realist's answer is incorrect. (In particular, Realist's answer does *not* follow from the dominance principle in that Bayes net game.) It is only by being sloppy in specifying the underlying Bayes net game that it appears that there is a conflict between the expected utility principle and the dominance principle.

Note also that no time variable occurs in our analysis of Newcomb's paradox. Concretely, nothing in our analysis using Bayes nets and associated games requires that $W$'s prediction occur before your choice. This lack of a time variable in our analysis means we can assign times to the events in our analysis any way we please, and the analysis still holds; the analysis is time-reversal invariant.

This invariance helps clarify the differences in the assumptions made by Realist and Fearful. The invariance means that both the formal statement of the paradox and its resolution are unchanged if the prediction occurs *after* your choice rather than (as is conventional) before your choice. In other words, $W$ could use data accumulated up to a time *after* you make your choice to 'retroactively predict' what your choice was. In the extreme case, the 'prediction' algorithm could even directly observe your choice. All of the mathematics introduced above concerning Bayes nets, and possible contradictions still holds.

In particular, in this time-reversed version of Newcomb's scenario, Fearful would be concerned that $W$ can *observe* his choice with high accuracy. (That's what it means to have $P(g \mid y)$ be a delta function whose argument is set by the value of $y$.) Formally, this is exactly the same as the concern of Fearful in the conventional Newcomb's scenario that $W$ can *predict* his choice with high accuracy.

In contrast, in the time-reversed version of Newcomb's scenario, Realist would believe that he can guarantee that his choice is independent of what $W$ says he chooses. (That's what it means to have $P(y \mid g)$ equal some $h(y)$, independent of $g$.) In essence, he assumes that you can completely hide your choice from $W$, so that $W$ can only guess randomly what choice you made. Formally, this assumption is exactly the same as the 'free will' belief of Realist in the conventional Newcomb's scenario that you can force *W*'s prediction to be independent of your choice.

In this time-reversed version of Newcomb's scenario, the differences between the assumptions of Realist and Fearful are far starker than in the conventional form of Newcomb's scenario, as is the fact that those assumptions are inconsistent. One could use this time-reversed version to try to argue in favor of one set of assumptions or the other (i.e., argue in favor of one Bayes net game or the other). Our intent instead is to clarify that Newcomb's question does not specify which Bayes net game is played, and therefore is not properly posed. As soon as one or the other of the Bayes net games is specified, then Newcomb's question has a unique, correct answer.

## 3 Conclusion

Newcomb's paradox has been so vexing that it has led some to resort to non-Bayesian probability theory in their attempt to understand it [22, 8], some to presume that payoff must somehow depend on your beliefs as well as what's under the boxes [11], and has even even led some to claim that quantum mechanics is crucial to understanding the paradox [13]. This is all in addition to work on the paradox based on now-discredited formulations of causality [7].

Our analysis shows that the resolution of Newcomb's paradox is in fact quite simple. Newcomb's paradox takes two incompatible interpretations of a question, with two different answers, and makes it seem as though they are the same interpretation. The lesson of Newcomb's paradox is just the ancient verity that one must carefully define all one's terms.

**Acknowledgements** We would like to thank Mark Wilber and Charlie Strauss for helpful comments.

|               | **Choose *AB*** | **Choose *B*** |
|---------------|-----------------|----------------|
| **Predict *AB*:** | 1000            | 0              |
| **Predict *B*:**  | 1,001,000       | 1,000,000      |

**Table 1:** The payoff to you for the four combinations of your choice and *W*'s prediction.

## APPENDIX

In the text, it is claimed that if for some $\alpha \in (1/2, 1]$, $W$ can ensure that $P(g \mid y) = \alpha \delta_{g,y} + (1 - \alpha)(1 - \delta_{g,y})$ for all $y$ such that $P(y) \neq 0$, and if in addition $P(y \mid g)$ is $g$-independent for all $g$ such that $P(g) \neq 0$, then in fact $P(y \mid g)$ is one of the two delta functions, $\delta_{y,AB}$ or $\delta_{y,B}$. This can be seen just by examining the general $2 \times 2$ table of values of $P(g, y)$ that is consistent with the first hypothesis, that $P(g \mid y) = \alpha \delta_{g,y} + (1 - \alpha)(1 - \delta_{g,y})$:

|          | $y = AB$ | $y = B$ |
|----------|----------|---------|
| $g = AB$: | $\alpha z_{AB}$ | $(1 - \alpha) z_B$ |
| $g = B$:  | $(1 - \alpha) z_{AB}$ | $\alpha z_B$ |

**Table 2:** The probabilities of the four combinations of your choice $y$ and $W$'s prediction $g$, given that $P(g \mid y) = \alpha \delta_{g,y} + (1 - \alpha)(1 - \delta_{g,y})$ for all $y$ with non-zero $P(y)$. Both $z_{AB}$ and $z_B$ are in $[0, 1]$, and normalization means that $z_{AB} + z_B = 1$.

If both $z_{AB} \neq 0$ and $z_B \neq 0$, then neither $P(g = AB)$ nor $P(g = B)$ equals zero. Under such a condition, the second hypothesis, that $P(y \mid g)$ is $g$-independent for all $g$ such that $P(g) \neq 0$, would mean that $P(y \mid g)$ is the same function of $y$ for both $g$'s. This in turn means that

$$\frac{\alpha z_{AB}}{(1 - \alpha) z_B} = \frac{(1 - \alpha) z_{AB}}{\alpha z_B} \qquad (3)$$

which is impossible given our bounds on $\alpha$. Accordingly, either $z_{AB}$ or $z_B$ equals zero. **QED**

# References

[1] Fudenberg, D. & Tirole, J. *Game Theory* (MIT Press, Cambridge, MA, 1991).

[2] Myerson, R. B. *Game theory: Analysis of Conflict* (Harvard University Press, 1991).

[3] Osborne, M. & Rubenstein, A. *A Course in Game Theory* (MIT Press, Cambridge, MA, 1994).

[4] Nozick, R. Newcomb's problem and two principles of choice. In *Essays in Honor of Carl G. Hempel*, 115 (Synthese: Dordrecht, the Netherland, 1969).

[5] Gardner, M. Mathematical games. *Scientific American* 102 (1974).

[6] Bar-Hillel, M. & Margalit, A. Newcomb's paradox revisited. *British Journal of Philosophy of Science* **23**, 295–304 (1972).

[7] Jacobi, N. Newcomb's paradox: a realist resolution. *Theory and Decision* **35**, 1–17 (1993).

[8] Hunter, D. & Richter, R. Counterfactuals and newcomb's paradox. *Synthese* **39**, 249–261 (1978).

[9] Campbell, R. & Lanning, S. *Paradoxes of Rationality and Cooperation: Prisoners' Dilemma and Newcomb's Problem* (University of British Columbia Press, 1985).

[10] Levi, I. A note on newcombmania. *Journal of Philosophy* **79**, 337–342 (1982).

[11] Geanakoplos, J. The hangman's paradox and newcomb's paradox as psychological games (1997). Yale Cowles Foundation paper, 1128.

[12] Collins, J. Newcomb's problem. In *International Encyclopedia of the Social and Behavioral Sciences* (Elsevier Science, 2001).

[13] Piotrowski, E. W. & Sladkowski, J. Quantum solution to the newcomb's paradox (2002). Http://ideas.repec.org/p/sla/eakjkl/10.html.

[14] Burgess, S. The newcomb problem: An unqualified resolution. *Synthese* **138**, 261–287 (2004).

[15] Benford, G., Book, D. & Newcomb, W. *Physical Review D* **2**, 263 (1970).

[16] Binder, P. Theories of almost everything. *Nature* **455**, 884–885 (2008).

[17] Wolpert, D. H. Physical limits of inference. *Physica D* **237**, 1257–1281 (2008). More recent version at http://arxiv.org/abs/0708.1362.

[18] Pearl, J. *Causality: Models, Reasoning and Inference* (Cambridge University Press, 2000).

[19] Koller, D. & Milch, B. Multi-agent influence diagrams for representing and solving games.

[20] Berger, J. M. *Statistical Decision theory and Bayesian Analysis* (Springer-Verlag, 1985).

[21] Starmer, C. Developments in non-expected utility theory: the hunt for a descriptive theory of choice under risk. *Journal of Economic Literature* **38**, 332–382 (2000).

[22] Gibbard, A. & Harper, W. Counterfactuals and two kinds of expected utility. In *Foundations and applications of decision theory* (D. Reidel Publishing, 1978).